\documentclass{article}
\usepackage{graphicx} 
\usepackage{biblatex}
\usepackage{authblk}
\usepackage{parskip}
\usepackage{url}
\providecommand{\keywords}[1]
{
  \small	
  \textbf{\textit{Keywords---}} #1
}

\title{Opening Musical Creativity? \\ Embedded Ideologies in Generative–AI Music Systems}
\author{Liam Pram}
\author{Fabio Morreale}
\affil{Waipapa Taumata Rau (The University of Auckland)\\Aotearoa (New Zealand)}
\date{August 2025}

\begin{document}
\maketitle

\begin{abstract}
    AI systems for music generation are increasingly common and easy to use, granting people without any musical background the ability to create music. Because of this, generative–AI has been marketed and celebrated as a means of democratizing music making. However, inclusivity often functions as marketable rhetoric rather than a genuine guiding principle in these industry settings. In this paper, we look at four generative–AI music making systems available to the public as of mid–2025 (AIVA, Stable Audio, Suno, and Udio) and track how they are rhetoricized by their developers, and received by users. Our aim is to investigate ideologies that are driving the early–stage development and adoption of generative–AI in music making, with a particular focus on democratization. A combination of autoethnography and digital ethnography is used to examine patterns and incongruities in rhetoric when positioned against product functionality. The results are then collated to develop a nuanced, contextual discussion. The shared ideology we map between producers and consumers is individualist, globalist, techno–liberal, and ethically evasive. It is a ‘total ideology’ which obfuscates individual responsibility, and through which the nature of music and musical practice is transfigured to suit generative outcomes.
\end{abstract}

\keywords{Artificial Intelligence; Democratization; Generative–AI; Ideology; Music }

\section{Introduction}
\textit{Unlocking musical creativity} has been the dominant marketing strategy of music technology companies \cite{McPhersonAndrew2019MIfN,morreale2020anime}. Their narrative is that people often have musical ideas they feel compelled to share, but these ideas remain untapped as they cannot spend thousands of hours learning a new instrument. With their purported solutions, anyone can \textit{finally} express their musical ideas with little difficulty. A similar strategy, which casts \textit{musicianship} and \textit{difficulty of use} as barriers to entry \cite{strum2024musaicology,morreale2020anime}, seems to be the strategy used by generative–AI (GenAI) companies to convince the public about the usefulness and the just grounding of their products.

In this article, we aim to critically examine the ideological foundations of mainstream GenAI music companies by investigating how AI–music researchers, developers, and entrepreneurs discuss the accessibility of their products. Specifically, we performed a systematic investigation into four GenAI music systems - Suno, Udio, AIVA, and Stable Audio - to identify what ideological positions are shared across the field. In order to support this analysis, we conducted a mixed–method investigation. We adopted a combination of autoethnography and digital ethnography to gather data from our own experience using these systems, from the companies themselves, and from their user bases. We then conducted an inductive thematic analysis on such data, identifying and articulating the dominant themes that reveal these companies’ ideological positions as well as patterns and incongruities in rhetoric when positioned against product functionality.

This paper contributes to the growing body of critical research into the use of generative AI in music and the arts \cite{barenboim2024exploring,ben2021music,bindi2023ai,krol2025exploring,MehtaAtharva2025MfAR}. As a technology with the potential to \textit{disrupt} creative practices, there are many concerns about whether its growing presence in the creative industries represents an imminent replacement of practitioners, or a continuation of digitalization, and processes of displacement in the ecosystem \cite{erickson2024ai}. Additionally, as an \textit{emerging} space, speculation from both developers and entrepreneurs as to the future of AI in music will undoubtedly influence the practical design directions the technology takes.

By critically studying the ideological positions held by various stakeholders involved with GenAI music systems at this early stage of development, we aim to help practitioners and scholars prepare for future impacts. As noted by critic Jan Soeffner, we cannot reflexively outpace disruption; instead, we are best off acting “always ahead of time, in order not to be surpassed by it” \cite[p.~3]{SoeffnerJan2025AoD}. Furthermore, this study offers a unique critical framing for the discussion of ideology in GenAI music and advances academic interest in the political and sociological underpinnings of this industry.

\section{Background}
In this section scholarly context is provided regarding relevant aspects of generative–AI music, democratization, and ideology.

\subsubsection{Generative–AI Music}
Generative-AI (GenAI) has been defined by Brindha and colleagues as “computational techniques capable of generating new and important content from training data, such as text, images, and audio” \cite[p.~2]{brindha2025introduction}, and by Feuerriegel and colleagues as “computational techniques that are capable of generating seemingly new, meaningful content such as text, images, or audio from training data” \cite[p.~111]{feuerriegel2024generative}. In these definitions, there is a focus on content which is \textit{new}, and \textit{meaningful}. Additionally, García–Peñalvo and Vázquez–Ingelmo note there is an understanding that the meaningfulness of the material hinges on its ability to be “\textit{engaged with} and \textit{consumed}” \cite[pp.~7–8]{garcia2023what}. A full understanding or identification of GenAI may rest then on both the production and reception of content.

Historically, formal methods have been used in music composition to “go beyond the known, and help generate something that cannot be directly envisioned” given the limited human capacities for mental modeling \cite[p.~53]{dahlstedt2018action}. Examples include the musical dice games of the later Eighteenth century, John Cage’s conceptions of chance operations, and Iannis Xenakis’s stochastic compositions \cite{braguinski2022mathematical,collins2018origins}.\\
Recent developments in AI music generation have extended both the technical quality and novel potential of music generations, rendering outcomes that may be indistinguishable from human craftsmanship \cite{morreale2025reductive,feuerriegel2024generative}. GenAI music may then represent an advancement along this lineage in which technology\footnote{Here we attribute “technology” a broad understanding, as in “the use of scientific knowledge for practical purposes or applications.”} has been used to reach beyond the human.

As AI finds meaningful engagement from both professional and amateur musicians \cite{ben2021music,banar2023tool}, it functions across a spectrum of creative positions; from a passive assistant to a co–creative partner or even a fully autonomous creator \cite{TigreMouraFrancisco2021AibB}. At the same time, the public's understanding of AI in music is dominated by the most popular GenAI systems. These systems are not particularly situated within any particular artistic practice, - for example text–to–music waveform generators, which dominate the ecosystem, do not serve highly specific artistic functions - but aim to be \textit{everything machines} for the public in a musical context \cite{bindi2023ai}. Despite the diverse array of potential uses for AI in music production and composition \cite{boateng2025ai}, public engagement with GenAI–music systems tends towards the restricted range of human–AI co–creative relationships facilitated by complete audio outcomes derived from text prompts.

Bypassing arguments about the location of intelligence in AI systems, the various creative roles held by AI transmit intelligent behavior, given to us by our culture, and from which we are shaped \cite{magnusson2019sonic}. While many researchers have highlighted the potential for supportive, reciprocal, creative relationships between artists and AI, historical lines have been traced between this leap in automation and other historic turning points, such as the industrial revolution, evoking related ethical and sociological concerns \cite{MorrealeFabio2021WDtB,de2025towards,fui2023generative,jiang2023ai}. Critics also note a potential lack of authenticity and intentional agency in AI content, which may appear generic as it “fails to mirror the internal motivation found to [...] drive human creativity” \cite[p.~138]{TigreMouraFrancisco2021AibB}. This perception has implications for both the reception of and engagement with AI in music, particularly for the general public, who might be suspicious of autonomously produced artworks. Ansani and colleagues \cite{ansani2025ai} indeed found that perceived involvement of AI in the creation of music, led participants to unfavorably critique the aesthetic and creative value of the music, as well as misjudge ‘objective’ issues with the performance when evaluated against works with perceived human involvement.

Strum and colleagues \cite[p.~13]{strum2024musaicology} sought to further contextualize the potential impacts of AI music on artists by suggesting a “musAIcking” to account for the ways in which musicians internalize the knowledge frameworks embedded in new technologies, which may have “a profound aesthetic influence on the music” that is composed, and on future directions of musical development \cite[p.~50]{dahlstedt2018action}. The underlying suggestion is that new critical tools may need to be developed to grapple with AI in music \cite{strum2024musaicology}, instead of recycling those used in previous musicological studies. This article engages with this call by suggesting a theoretical framing to critically analyze such tools.

\subsection{Democratization}
Advances affecting the potential applications of GenAI in music have come hand–in–hand with a significant increase in the ease–of–use of many GenAI–music interfaces. As potential profits rest on the development of a broad user base, many of the biggest investments in the sector have been developing GenAI–music systems for general users. As such, accessibility for non–musicians has become a key concern for developers, and as in other sectors being \textit{disrupted} by AI technologies, \textit{democratization} has become a buzzword \cite{eff2023ai,louder2024how,marr2023generative}.

Common use of the term “Democratization,” particularly in relation to technology, describes “the action of making something accessible to everyone” \cite{democratization2010}. Composer and music technologist Tae Hong Park describes democratization in music as a phenomenon propagated “through affordable, sophisticated, and cost–effective technologies enabling non–experts to engage in music–making” \cite[p.~18]{park2017instrument}. Although democratization may imply magnanimity, a focus on \textit{access and ease} ignores the specific desires of users \cite{morreale2020anime}. Additionally, the large–scale GenAI systems that purport to democratize musical processes hold a significant Western bias \cite{MehtaAtharva2025MfAR}. AI becomes a solution driven by an invisible demand, “[reaching] for answers before the questions have been fully asked” \cite[p.~3]{morreale2020anime}.

Sociologist Nancy Weiss Hanrahan critiques the use of democratization in which “democracy is conceived as a technological rather than a historical achievement” \cite[p.~207]{hanrahan2022music}. She argues that the evocation of democracy in relation to developing music technology ignores the contingent nature of the term and turns participation into a binary where access is necessarily good. Democratization is aligned with modern “democratic” institutions that operate under libertarian normative principles, centering individualist concerns. Additionally, she argues we need to reconsider the foundational reasoning behind “why music matters”, and how it can offer something positive against which we can offer critique, aligning with Harkins and Prior \cite{harkins2022dis} in questioning whether democracy is a valid or appropriate criterion for music in the first place.

Sturm and colleagues \cite{strum2024musaicology} find democratization a recurrent theme in the literature around AI and creativity, and a position utilized in the commercial sector due to its “[appeal] to a massively scalable business model,” which offers greater resilience than highly targeted ones. This lends credence to critical appraisals of democratization’s potential altruism. The drive to expand the population of music makers is likely a shallow pursuit of growth over meaningful engagement, and democratization could represent a “form of technologically–mediated deskilling,” reforming the kinds of musical tasks that are left to the user \cite[p.~7]{strum2024musaicology}. These so-called ‘democratizing’ systems can operate in extractive and reductive ways, harvesting users’ creative input while withholding meaningful insight into their mechanisms, which stay largely opaque \cite{strum2024musaicology,masu2021composing}.

By exploring the context of democratization, we can see how its explicit use by technologists may expose a great deal about their motivations and underlying ideologies.

\subsection{Ideology}
As our analysis is framed through ideology, we begin by clarifying the concept. Additionally, a brief discussion of relevant established ideologies that will be discussed later in the article is presented.

Social theorist Manfred B. Steger defines ideology as a “comprehensive belief [system] comprised of patterned ideas and claims to truth” \cite[p.~217]{steger2013political}. As these systems create links between belief and practice, they attempt to fix definitions of core concepts in what political scientist Micheal Freeden terms ‘decontestation’ \cite[pp.~60–91]{freeden1996ideologies}. Those engaging in ideology may be implicitly or explicitly directed by these belief structures. However, ideologies are not \textit{necessarily} coercive, and can represent all kinds of power dynamics, including lateral social interactions and developments via discourse which are non–hierarchical. As Teun A. Van Dijk \cite{vandijk2013ideology} argues, there are no personal ideologies; ideologies are socially shared belief systems. In this way, he asserts that ideology is “largely acquired, expressed, and reproduced by discourse” \cite[p.~176]{vandijk2013ideology}. This social understanding is key to our investigation of ideological expressions across the AI ecosystem.

Many of the ideological positions ascribed to the development of AI can be understood against the cultural backdrops of \textit{neoliberalism} and \textit{techno–liberalism} \cite{atanasoski2019surrogate,morreale2023data,bourne2019ai,chatterjee2024empire}. Neoliberalism constitutes social, cultural, and political–economic forces that center unregulated private competition as the optimal means of developing “a truly free social world” \cite[p.~2]{Wilson2018N}. Neoliberal \textit{discourse} is identified as a tool used to forward the unregulated expansion of AI, which aligns itself with core neoliberal processes like individualization and perpetual consumption. Legal scholar \cite{chatterjee2024empire} observes that under neoliberalism, AI operates at both the granular level of the self, where it offers affordances to the requests of the individual, and the global level, where collections of selves constitute both the data used to construct AI systems, and the target consumers.

Neoliberal allegiance with and deployment of AI technologies are often discussed under the sub–paradigm \textit{techno–liberalism}, an ideology which manifests an “abiding faith in the power of technology and managerial technique to solve problems and help preserve liberty” \cite[p.~103]{baer2017democratic}. Sociologist Nikolas Rose argues that such political ideologies render institutions controllable, “intelligible, and calculable” in technical processes via data \cite[p.~197]{rose1999powers}. As these institutions develop around or construct communities, they take hold of “active practices of self–management [...], identity construction, [...] personal ethics and collective allegiances” in what he terms \textit{government through community} \cite[p.~176]{rose1999powers}. In this way, subjects may then be indirectly oriented towards the ends of the managerial body. Chatterjee notes that this pattern of unconscious consumption “envisages an uncritical conformation of individualized and private consumers with the neoliberal market logic” \cite[p.~9]{chatterjee2024empire}. In executing these processes, techno–liberal ideologies deploy a \textit{Techno–solutionist} logic in which all issues can be reduced to technologically solvable problems, and where issues themselves may be manifested expressly for the purpose of justifying development \cite{morreale2020anime}. The attempt to democratize music via AI technologies may represent such a case.

\section{Methodology}
\subsection{Process}
This research is structured in three stages: an autoethnographic engagement with GenAI systems; a digital ethnography exploring the selected companies and available rhetoric from their representatives; and a digital ethnography in which collected data from the selected online communities developed around the user bases for each company. By undertaking this mixed methodology, we were able to compare and discuss results derived from three different loci of attention; ourselves, the companies, and the user bases, attending to gaps that may emerge from a siloed focus on each.
Finally, intermittent thematic analysis was performed on the data collected in each phase of the research to discover prevalent themes for discussion. A visual representation of the methodological form is presented in Figure 1.

\begin{figure}
    \centering
    \includegraphics[width=0.8\linewidth]{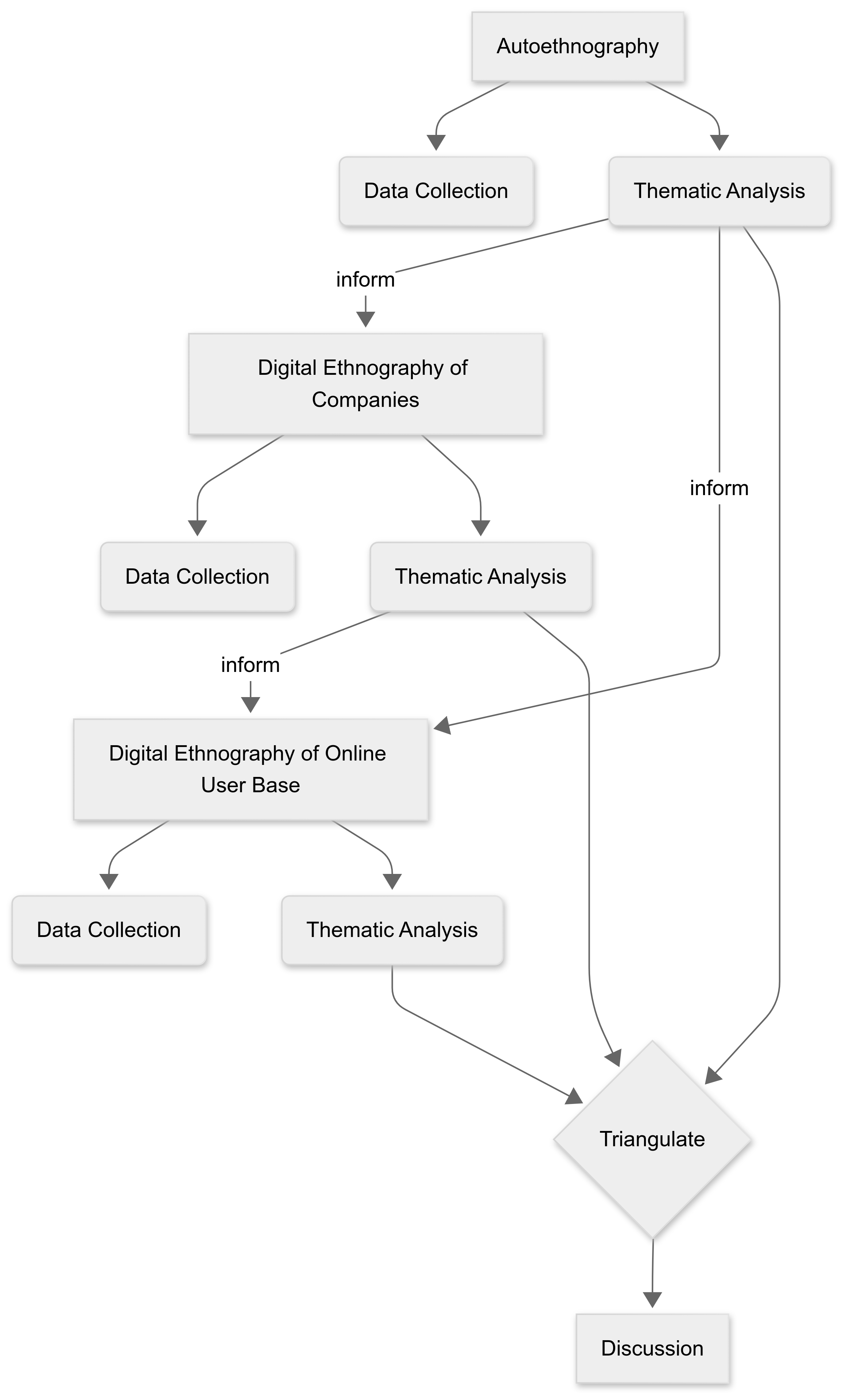}
    \caption{A flowchart representing the methodological form of the study}
    \label{fig:1}
\end{figure}

\subsection{Selection of GenAI Music Systems}
GenAI music systems were selected using nine criteria synthesized from the objectives of the study:
\begin{enumerate}
    \item Uses generative AI
    \item Not released after May 31st 2024
    \item Currently active
    \item Makes explicit claims around democratization
    \item Aimed at the general public
    \item Uses text, audio, or symbolic prompts
    \item Focuses on complete music generation
    \item For non–specific use
    \item Has an active online community
\end{enumerate}

\textit{Not released after May 31st 2024} and \textit{Currently active}\\
A temporal limitation was imposed on the selection process to account for the rapid state of development in the field. Systems which were produced or made available after May 31st 2024 were not considered. The chosen AI music systems needed to be currently active to reflect the state of the industry. A total of 35 generative AI music systems were identified that met the first three criteria. These systems were then filtered through the subsequent six criteria.

\textit{Makes explicit claims around democratization}\\
An initial survey to find claims of creative democratization was used to establish whether these systems were engaged in the ideological space being mapped by this study. 

\textit{Aimed at the general public}\\
As democratization is a core subject of the present investigation, it is important that these systems were available to, and aimed at, the general public. Systems that required coding literacy to be used meaningfully\footnote{As was the case with MusicLM, MusicGen, AudioGPT, ChatMuscian, InstrumentGen, HarmonAI, JukeBox, and MuseNet.} were excluded, as this posed a barrier to broad participation.

\textit{Uses text, audio, or symbolic prompts }\\
Systems that erected a barrier to \textit{personalized} input, and thus compromised the expressive potential for specific, individual creativity, were avoided. If the creative potential of the systems was compromised from the outset, we would hypothesize limited meaningful engagement from users. As a result, we worked with systems largely using text, audio, or symbolic prompts, avoiding more restricted systems, such as those which only use categorical selection\footnote{Users can select, for instance, tempo, genre, and instrumentation - as seen in Boomy, Soundraw, Splash, Soundful, MusicTGA, and MusicStar.}.

\textit{Focuses on complete music generation} and \textit{For non–specific use}\\
Systems that focus on particular musical elements, or adjacent sound craft such as instrument building, production, or deepfakes were not included in this study. Though these may be legitimate forms of music composition, we argue that such systems\footnote{InstrumentGen, Musicfy.} manipulate musical elements, and are not engaged with what would be \textit{generally} understood by the public as complete ‘music generation.’ Systems which appealed to a highly specific user base or function\footnote{As with DAACI, Emergent Drums 2, Splash, Infinite Album, Musicfy, and Brain.fm.} were also avoided to maintain general relevance to a public user base.

\textit{Has an active online community}\\
This research looks partially at user experiences to see how ideology manifests itself through GenAI–music systems into a broad public sphere. It was thus important that the subjects of the study had active online communities, engaged in using and discussing the systems. Some systems did not meet this criteria because they lacked an online community,\footnote{Junia, Mubert, Soundverse, Hydra, InstrumentGen, ChatMusician, AudioGPT, MusicGen, MusicFX, MusicLM.} or had communities which appeared to be currently inactive.\footnote{Riffusion, Cassette AI, MusicLang.} We conducted our community research primarily on Discord.\footnote{https://discord.com/}

Four systems fulfilled all nine criteria; AIVA,\footnote{https://www.aiva.ai/} Stable Audio,\footnote{https://stableaudio.com/} Suno,\footnote{https://suno.com/} and Udio.\footnote{https://www.udio.com/}

\subsection{Autoethnography}
An autoethnography was undertaken to establish emergent themes in response to to our first–person experience with using these systems. As creative empowerment is marketed as one of the democratizing benefits of these systems, we wanted to test this promise by exploring a range of goals based around aspects of creativity. Anna Jordanous’s \cite{jordanous2012standardised} 14 point “working understanding of creativity” was used to motivate these autoethnographic goals and procedures. Four goals were pursued with each system to explore a variety of musical ambitions while falling within the practical scope of the research. Each goal was pursued as far as viable in each system, and then attempted in the next to ensure continuity of approach and outlook when exploring each goal. Reflections were logged during each generation procedure. 

The notes then underwent inductive thematic analysis as outlined in Section 3.5. The entire data corpus was analyzed across the different platforms to identify universal themes that could help direct what secondary data would be collected from the companies and developers in the next stage of the study, beyond that pertaining to democratization.

\subsection{Digital Ethnographies}
\subsubsection{Companies}
A digital ethnography was then undertaken to collect and analyze secondary data from the companies behind each AI music system. Targeted searches were used to collect secondary data sources that were publicly available from the companies and key personnel (CEO’s, lead developers, etc.). For interviews, podcasts, and panel discussions (audio or video), an automatic speech recognition system was used to transcribe the audio into text for analysis.

\subsubsection{Online Communities}
Thematic analysis (see \textit{Section 3.5}) of secondary data  from the companies was used to generate a group of key themes which would be our focus while gathering data from the online communities:

\begin{figure}
    \centering
    \includegraphics[width=0.9\linewidth]{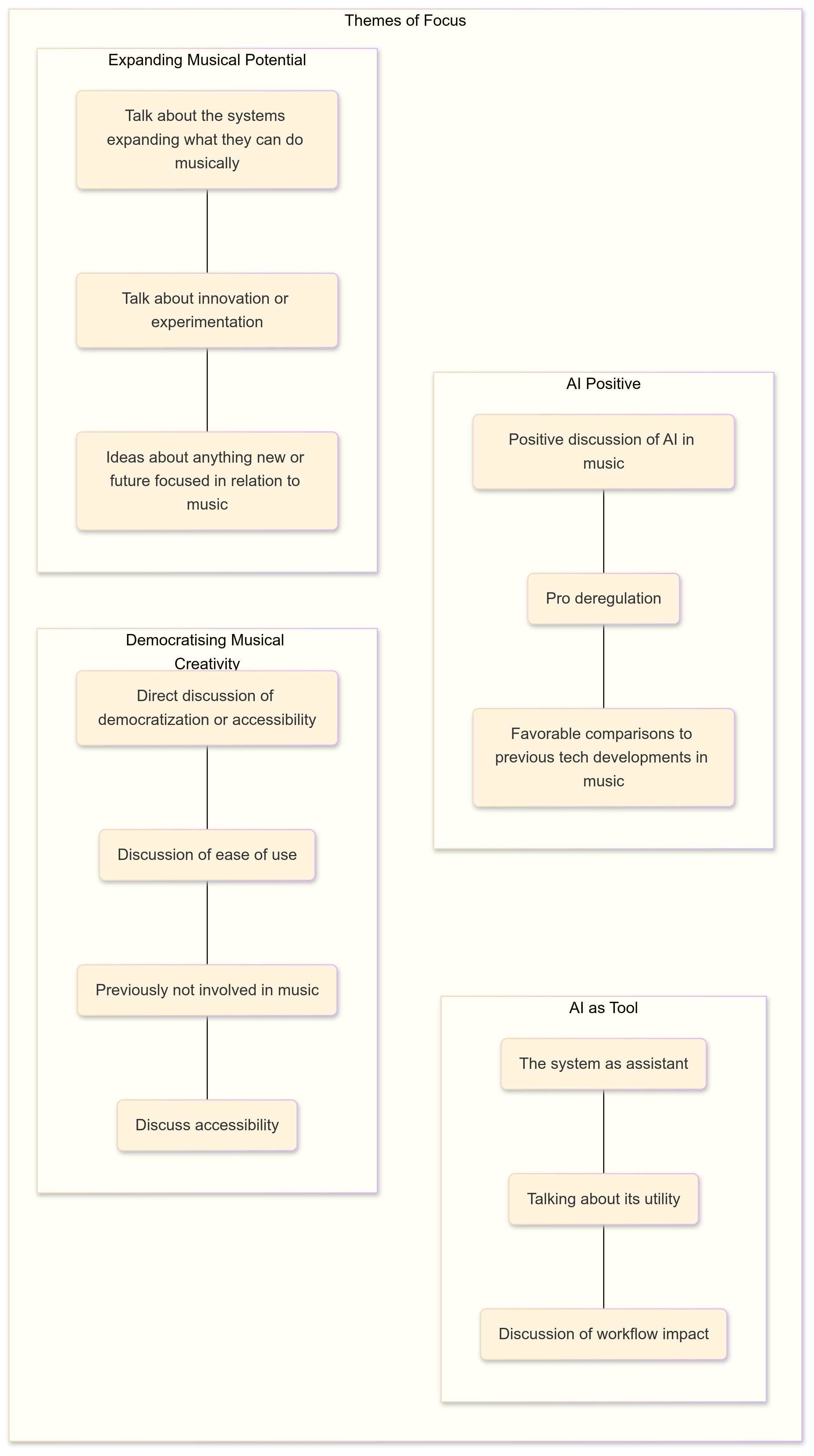}
    \caption{Themes of focus for companies' ethnography.}
    \label{fig:2}
\end{figure}

The online user communities of all four GenAI music systems examined in this study were based on Discord.\footnote{There was the inclusion of some material from Reddit, but Discord was by far the main repository from which data was collected.} All available data was collected from the start of 2024 to the end of August for AIVA and Stable Audio. As the Suno and Udio communities were highly active,\footnote{At least one post per minute.}. Targeted searches proved a more effective strategy to deal with the larger data corpus. Further temporal delimitation was required for some search results, these are noted in italics. The terms used for targeted searches were:
\begin{itemize}
    \item "democratise"/"democratize"
    \item "easy to use"
    \item "musical creativity"
    \item "creativity" \textit{1 Jun – 31 Aug}
    \item "musicianship"
    \item "innovation"
    \item "originality" \textit{1 Aug – 31 Aug}
    \item "future of music"
    \item "experimentation" \textit{1 Jun – 31 Aug}
    \item "new music"
    \item "AI as tool"
    \item "assists"
    \item "collaboration"
    \item “workflow”
\end{itemize}
As we discovered relevant data which did not directly involve these keywords it was also collected. The collected data also underwent thematic analysis (see next section).

\subsection{Thematic analysis}
The thematic analysis was undertaken based on procedures outlined by Braun and Clarke \cite{braun2006using} and was used to analyze both the autoethnography and digital ethnographies of secondary data (companies and online communities). The analysis followed a constructivist epistemology, focusing on the role of active knowledge construction and socio–cultural influences of the data corpus. The research attends to individual perspectives without positioning them as its primary subject, and without asserting the truth of any particular one. Instead, we focus on what ideas are present in and supporting public-facing GenAI–music system development, and from what sociocultural and structural positions they emerge, investigating what is beneath the available discourse. Driven by the research question and interstitial findings throughout the study, inductive and theoretical approaches are used while remaining open to any themes that might ‘emerge’ from the data. 

The persistent and problematic presence of democratization in discussions around generative–AI in music was a motivating force for this research, subsequently directing the subjects and data that initiated our process. As the study continued, other subjects and themes directly or tangentially connected to democratization emerged. The research remained flexible to these findings, with democratization acting as a core concern and springboard, but not the sole subject of interest.

\section{Results}
The results of each section of the study are grouped into the themes that were individuated during the thematic analysis. 

\subsection{Autoethnography}
\subsubsection{Control and Renegotiation}
We found that many of the systems \textit{directed} our creative engagement in ways that were both enabling and inhibiting, noting that “any kind of [precise] control felt difficult.” When outputs failed to adhere to our expectations, we were either forced to submit to, renegotiate, or exit the interaction. During a procedure, we noted, “[it] just seemed to have recognized the instrumentation, and ignored my prompt.” This interruption to the inherent spontaneity of the interactions fractures engagement, and threatens meaning formation.

Certain creative goals felt unachievable due to perceived limitations in the systems. We noted issues trying to explore “alternative ways of thinking about time, harmony, and timbre (playing technique).” As we discovered these boundaries, our ambitions were regulated accordingly. By reducing the space between ideas and outcomes these systems demonstrate promising characteristics to springboard inspiration, particularly as part of a larger compositional process in which the generated outputs are not the final work. 

As these systems encourage continuous probing, we identified experimentation as a de facto means of engagement, noting that “because things are unclear, you need to be constantly probing[...] like exploring a cave in the dark.” When framing the interactions as experiments, errors that would otherwise be deemed to be failures are permissible, and valuable cross–platform knowledge was internalized through these probing procedures. 

\subsubsection{Normalization and Commodification}
These systems appeared to \textit{normalize} outcomes. During the autoethnographic engagement with the systems we noted that “the system seems to be designed with a particular kind of music in mind,” in that various attempts to render specific goals were driven in other directions. Systems that were rules-based or appeared to re–write prompts \textit{behind–the–scenes}\footnote{This is speculation based on our experiences.} had greater conventional tendencies and genre was often positioned as a basic musical element (ie. suggested by the systems at the prompting stage).

When our control over the system was mitigated, we found the process seemed to be taking place \textit{regardless} of our involvement, instead of as an \textit{extension} of our intentions. We noted that in one system interaction, it felt as though the system did not “want to allow the user much creative control. Perhaps they are afraid of generating 'failures' that may result from experimentation.” In this way, we feel the user is de–centered in exchanges with the system. Rather than feeling creativity empowered, the user may labor towards their goal in a combative relation with the system. Additionally, as systems streamline the process between production and publication, \textit{completion} is emphasized over \textit{process}.

\subsection{Digital Ethnographic of Companies}
\subsubsection{Democratizing Musical Creativity}
Representatives of all the systems reviewed made some claims to democratization. Focused on the general public, imagination and taste were cast as entry points into musical creativity, allowing anyone to express themselves musically without musicianship. “No instrument needed, just imagination. From your mind to music" \cite{suno.n.d.suno}.

Creativity is positioned as \textit{dormant}, the system acting as a “super–powered instrument that \textit{amplifies}” \cite{udio2024introducing}, \textit{augments}, but does not \textit{replace} human creativity, to “empower music enthusiasts and creative professionals to generate new content” \cite{stability2023stable}. Suno CEO Mikey Shulman suggests that music making is an innate part of being human, from which “we kind of get acculturated" \cite{wilson2024AI}. AI is offered as a technical means to “rediscover” this desire from which we have become disconnected \cite{shulman2024suno}. 

\subsubsection{Individualization and Humanness}
Company representatives argued that innovations in AI music will blur the boundaries between creation and consumption into unconstrained novel spaces that are yet to be imagined. Specifically, they give the example of users creating the music they desire for their own consumption. AIVA CEO Pierre Barreau says this may look like “a personalized life soundtrack for each and every individual, based on their story and their personality” \cite{chivot2019barreau,MorrealeFabio2021WDtB}, while Shulman says of the technology; “what we have done is we have opened up a new method of interaction [...] beginning to blend or blur the line between creation and consumption” \cite{newcomer2024future}.

While claims are made that AI extracts “high–level features of what makes human music human music” \cite{kravitz2017pierre}, representatives suggest AI \textit{understands} music in a manner significantly different from humans, potentially learning “some abstraction of music theory that is different from what we learn” \cite{todd2023stability}, introducing novel musical potentials.

\subsubsection{Techno–liberal Perspectives}
We find a \textit{hands off} deregulatory approach to technological development, centering individual ethical responsibility. Zach Evans from Stable Audio argues “[regulation is] what brought down Napster, that's what brought down all the fun things in music they blow up” \cite{weights2022}. More concretely, direct ties have been established between lead investors in Udio, and a submission to the US Copyright Office “arguing that training AI on copyrighted materials should be lawful in the States [...] and doesn’t amount to theft of intellectual property” \cite{tencer2024new}.

The development of AI is framed as inevitable but unthreatening, and historical antecedents are commonly used to dissuade fears. An Udio representative argues: “virtually every new technological development in music has initially been greeted with apprehension, but has ultimately proven to be a boon for artists, record companies, music publishers, technologists, and the public at large” \cite{udio2024today}.

\subsection{Digital Ethnography of User Base}
\subsubsection{Democratization and Dormant Musical Ability}
The communities we reviewed tended to embrace AI as a democratizing force allowing users to overcome material limitations. One Udio user says "I believe in the power of AI to democratize content creation and empower those who don't have the resources to create something themselves.” Some users feel this democratization unlocks a dormant musical ability, a Suno user commenting it “feels like it's unlocked something in me that was laying dormant all these years because I never connected with an instrument,” while another says "it sparked my creativity for sure [...] it was always in my head but [I] couldn't get it on paper."

\subsubsection{Augmentation and Assistance}
Community members often framed the systems as tools or instruments, \textit{augmenting} and \textit{assisting} but not \textit{replacing} human involvement in the process. A Suno user argues “ultimately, you are in control, and the tool is executing your vision." Integrating the technology into existing workflows, users expedite established practices, as explained by a Stable Audio user: “Instead of wasting hours searching for the perfect sample. I just describe it to the AI and get it.”
In its assistive capacity, AI may be framed as a collaborator, and compared directly with human capabilities. A Suno user describes the system as “a collaborator who will take my direction, "while another feels "it is mimicking what it hears, just like human musicians do.”

\subsubsection{Innovation and Creativity}
Many users feel AI will improve and innovate the musical landscape, a feeling expressed by three Suno users: "It's like suno is imagining the future of music”; “I mean AI is the future of music”; “I think I've created a new music genre." Within this context, creativity is framed as a goal or measure for users. Regularly discussed both in terms of the originality of outcomes, and the level creative control available to the user, there appears to be some conflation between creativity and novelty, with randomness in the system viewed as both creatively essential, and a barrier to control. The exploration of ‘innovative’ or ‘explorative’ methods in the community often hinge on genre hybridity. An Udio user argues the future of music will be “trying to fuse different music genres,” while another says they “have often thought of mixing different vocal styles with different instrumentation styles.” 

\subsubsection{Techno–Inevitability}
Similar to what we saw from system representatives, users feel AI is an inevitable part of music’s future. Two Suno users argue it’s “not like we can put the cat back in the bag now [...] I don't know why people fear AI so much.” Seen as a natural progression from older technologies, AI is also framed as a \textit{new kind of instrument}, which has as of yet unreleased potential. Two Udio users argue; “Like those who refuse to use samples, or synths etc saying it's not ‘real’ music”; "many people are against AI. they are stuck in the past and will be forgotten like all the horse and buggy riders.”

\section{Discussions}
\subsection{Ideological Implications of Democratization}
Considering these results through the lens of ideology, we find several ‘claims to truth’ are made about AI in music with a central claim being that it enables \textit{independence}.. The idea of accessible creative independence is positioned as an improvement to the current state of musical engagement. Typical ways of music making are deemed too difficult, too multifaceted, too dependent on community engagement and infrastructure for the average person to access. An Udio user explains “many of us were having the opportunity to be creative without having a studio... or knowledge of music.” As this new technology promises to be \textit{easy}, it is postulated that any non–musician may enter into music making without commitment.

A useful lens to understand this phenomenon is Michel Foucault’s concept of \textit{governmentality}. He argues that, through a complex set of mechanisms aimed at guiding the conduct and choices of a population toward desired outcomes, forms of self–governance become central to the liberal model of the active citizen, operating both within and beyond traditional positions of authority (e.g., the state) \cite{foucault1991foucault,haughton2013spaces}. Foundational democratic notions of liberty are then challenged within this complex, in that the pursuit of liberty plays into processes of governmentality. As Foucault states, “power is exercised only over free subjects, and only insofar as they are free” \cite[p.~221]{foucault1982beyond}. The lens of ‘neoliberal governmentality’ asserts that the promise of a better and easier everyday life delivered through technical means decenters power, and exerts it over subjects who “are \textit{produced} as ‘free’ [and thus] are able and willing to consume as private individuals” \cite[p.~9]{chatterjee2024empire}.

As notions of democratizing music via GenAI rest on independence and creative empowerment, it is clear how the technology may be situated within models of governmentality, in acts of quiet coercion, particularly when the growth of access is privileged over depth of engagement. Suno declare they are “building a future where \textit{anyone} can make great music[...] No instrument needed, just imagination” \cite{suno.n.d.suno}. The barriers may have been lowered, but open discussions around \textit{what is great music} appear to be less of a priority.

Central to these processes of governmentality is an individualist ideology aimed at “guaranteeing each individual the possibility of the free development of their own personality” \cite[pp.~58–59]{gentile2013total}. Music is marketed to the free–individual as fundamental yet marginalized, an essential space that, Schulman insists, “most people interact with [...] only passively" \cite{oracle2025suno}.\footnote{Moving from passive to active consumption is something we see discussed elsewhere concerning leveraging AI applications in the arts towards capital \cite[p.~3]{hong2025leveraging}.} Engagement is then framed as a necessary \textit{choice}, and the universality of music is used to argue for widespread adoption. As the individual audience is transformed into a collective \textit{everybody}, appeals are made to ‘common’ experiences, with the technology presented as both a benevolent gift from a technocratic authority and as something emanating from the will of society at large, what Stable Audio terms “AI for the people, by the people." We see the varying degrees to which these companies rhetoricize democratization reflected in the community engagement with the concept,\footnote{Though its minor presence or absence in the Stable Audio and AIVA communities will be impacted by a general lack of community engagement noted previously.} evidence of successful ideological dissemination, regardless of intention.

Complementary to the centralization of the individual, we observed discussions around \textit{augmented AI} in which the technology should assist or extend human ability, not replace it. Often referred to as an \textit{assistant} or \textit{tool}, the ability of GenAI to spark or rejuvenate musicians' relationships with composition, and serve as inspiration for users engaged in “creative search” \cite[p.~515]{ben2021music} is highlighted in this conception. As one Suno user puts it, “generative AI is a good blueprinting tool [to] help with the creative process.” We see a preference for this relationship through the community discourse, consistent with previous findings \cite{mcclennan2025about,krol2025exploring} that musicians were more open to AI in music as a tool than as a collaborator, privileging control over levels of agency. When users conceptualize AI as an augmentation of the self, they further play into processes of self–governance by identifying with the technology, and complicating critical reflection.

While the ideology is distinctly individualist, it positions the individual within a \textit{globalist} framework, and the subject is expanded beyond society–members to the generality of humanity. This globalist framing shifts focus away from local contexts, both actual and conceptual, mitigating the need to engage with cultural specifics (Walker, 2012). By evoking a humanist position through their generalist rhetoric, system representatives conjure a theoretically benevolent AI that obscures any insidious means of profiteering. Focusing on the global application of this technology keeps the apparent benefits vague enough to be inscrutable, and as ‘friends’ to the needs of many, developers and companies then stand as a bridge between the public and art, industry and audience. We can understand this process through a Marxist lens in which ideology serves as a mystifying force used to develop a ‘false consciousness’ \cite{gentile2013total}, helping companies disguise the material impacts of these systems, while eschewing ethical concerns under the guise of technological inevitability. As illustrated by Schulman, one of the company's operating principles is “how do [we] more quickly pull forward the future of music that we envision and deploy capital toward that?" \cite{effron2024company}. The future is \textit{known}, it is AI, and so that is where they invest. This sense of inevitability subjugation to it is reflected in Suno’s user base. One user states, "well no one is really ready for this level of democratization of music, but we just need to adapt. Not worth it to protest or fight against it."

\subsection{Total Ideology}
Our results reveal that AI is commonly positioned as “a unified global process that unfolds in successive phases or stages of development” - what historian Emilio Gentile notes as a characteristic of the ‘total ideologies’ which coalesced following the cataclysm of the French Revolution \cite[p.~60]{gentile2013total}. The way in which this perspective manifests in the companies surveyed for this study is neatly summarized in this quote from AIVA CEO Pierre Barreau:
\begin{quote}
  Historically there have been other technical advances that have brought music–making to the masses, and these have not reduced human creativity. In fact, they have supercharged it \cite{luxndpierre}.  
\end{quote}

Framing GenAI in music as the inevitable outcome of prior technological developments enables companies to cast themselves, ostensibly altruistically, within a historical trajectory they position as beyond their influence. Within this narrative, AI-driven innovation in the music industry becomes \textit{inevitable}—accelerating existing practices while, more significantly, opening speculative and amorphous spaces that remain conveniently insulated from direct criticism. Framed as \textit{important} and \textit{emergent}, the best transformational potentials of AI are used to compel us to grant it free rein.

Total ideologies are holistic, putting “the individual at the service of society, rather than society being at the service of the individual” \cite[p.~58]{gentile2013total}. This would appear to exclude the distinctively individualist ideology we have mapped so far. However, the globalist neoliberal tendency to generalize that we have observed fits well with total ideologies' attempts to eliminate disagreement by defining its subjects as community members \cite{gentile2013total}. The democratization of music via AI is framed as a noble pursuit, a “watershed moment in technology [where we will be able to] expand the circle of creators, empower artists, and celebrate human creativity" \cite{udio2024today}. The companies and developers themselves then become subordinate to the holistic conception, in which their role as innovator swept along in this unified process legitimizes their actions beyond profiteering.

The total ideology is teleological in nature, moving towards human emancipation following the disruptions of modernity, and ever cognizant of a momentous collision between the old and the new \cite{gentile2013total}.
As the ideology we are mapping constructs AI as inevitable and decontested, the acceptance of the technology as part of the rhythms of historical change requires the global community undergo “internal renewal, intended to eradicate egoism from the human heart and thus create a new man animated by a collective sense of community” \cite[p.~116]{ozouf1989lhomme}. This ideology posits the “future [for] music where technology \textit{amplifies}, rather than replaces [...] human creativity” \cite{shulman2024suno} requires this ‘new man’ to constitute a new historic–social group oriented towards an imminent socioeconomic epoch, in the vein of the \textit{Fourth Industrial Revolution}, the \textit{Second Machine Age}, or \textit{Industry 4.0} \cite{atanasoski2019surrogate}\cite{hong2025leveraging}\cite{TigreMouraFrancisco2021AibB}.

Representatives of these systems draw equivalences between different creative acts, and between the past and the present. Zach Evans from Stability Audio says ``I mean, you can pick up a pencil, you can pick up a guitar,” and “historically, the big changes [...] in music very frequently come from advancements in technology" \cite{todd2023stability}. Additionally, comparisons are drawn to other technologies, Mikey Schulman suggesting AI will “let people make music for [their] friends and share it the same way you might do with an image” \cite{mignano2024mikeyschulman}. We can read this as an attempt to familiarize the audience with the unfamiliar, entrenching AI in music as part of a teleology that discounts the context of development, and fixates on a theoretical arrival.

Similarly, negotiated agency between the user and the system sometimes evolves to users drawing equivalences between technology and human:
\begin{quote}
    Tbh I don't view Udio as a tool or an instrument. I've come to respect it as an actual creative collaborator[...] I don't get why people say this isn't a form of creativity, though. It's quite literally just doing what we do.\footnote{Emphasis added.} – Udio user
\end{quote}

Composer and researcher Palle Dahlstedt argues we can relate to “music coming from algorithms in very much the same way as we relate to music coming from fellow musicians,” drawing parallels between thought and algorithms, which in a musical context may both result in similar concrete sonic outcomes, and thus illicit a similar empathetic dialogue between agents, regardless of their humanness \cite[p.~44]{dahlstedt2018action}.

As an assistive role for GenAI in music is emphasized by authorities and users, anthropomorphization is employed as a metaphorical extension of this assistive position, for example CEO Pierre Barreau’s use of ‘she’ to refer to AIVA – “she learns to predict when notes should come next in those tracks" \cite{dassault2020compsing}. This use of anthropomorphization acts as a means of defanging the technology while hinting at its creative and emotional abilities. As noted by Bindi and colleagues, narratives \textit{naturalizing} AI  imbue them with a sense of autonomy which allows developers and researchers to \textit{discover} them as if they were \textit{already existing} \cite{bindi2023ai}. This anthropomorphization may then be an attempt to limit the threat posed by the technology as a sudden leap forward, and ingratiate it into the total ideological form as a part of the human collective.

Additionally, as “the human nature of the work embodied in [a] product” has a major effect on what customers are willing to pay for in AI services, the rhetoric we see emerging from these spaces has many motivators for imbuing these systems with humanness \cite{TubadjiAnnie2021Cpbi}.

\subsection{Techno–liberalism and Dormant Creativity}
\begin{quote}
  The basis of freedom is the ability to create and express yourself without someone saying you can't do that. – Udio user  
\end{quote}
This early stage of GenAI is generally framed by both companies and users as a \textit{sink} or \textit{swim} scenario. As one Udio user puts it; “You can't push toothpaste back into the tube. You have to accept reality and collaborate with it.” We found that historical precedence set by other technologies (e.g.,s sampling, digital editing) is used to legitimize GenAI in music. Evans says “economically[...] I can compare it to previous changes in music, like, you know, the rise of the DAW probably puts some studio engineers out of work. But I think that was still worth it for the, you know, growth in music” \cite{day2024zachevans}. Where these antecedent technologies were initially seen as disruptive, they have in turn revolutionized production practices and entered into the normative musical lexicon. GenAI is positioned as part of this \textit{inevitable} and \textit{unified} process of technological progression, and one which is likely to be beneficial and creativity iconoclastic.

The ideological position that emerges from this narrative can is techno–liberal, in that it puts faith in technology as the central means of liberal development, centering individual needs and concerns, and the leveling effects of market competition. Competition, as an aspect of the ‘decontested’ neoliberal market, is positioned as a “civilizing force” \cite[p.~136]{turner2008neo}, where AI’s role in the growth of music – its potential to expedite workflow, improve and innovate musical possibilities, and democratize music making – is used to foster development as part of “a natural, ‘spontaneous’ order of ‘civilised’ values and mutual cooperation, which sustain” individual focused conceptions of freedom and constant expansion \cite{turner2008neo,baer2017democratic}.

Techno–liberalism posits that technological development will initiate a new phase of human emancipation, one which Atanasoski and Vora \cite{atanasoski2019surrogate} observe erases identity in an imagined era which is post race, post gender, and post labor. We can observe this depoliticizing mentality reflected in this quote from a Stable Audio user; ``Technology is neutral. It just depends on how you use it." By automating undesirable forms of labor, those “performed by racialized, gendered, and colonized workers in the past” \cite[p.~4]{atanasoski2019surrogate} the techno–liberal ideology ‘liberates human potential’ as a means of furthering capital interests, rendering invisible those consequences which are unappealing. The positioning of GenAI–music along this lineage invites scrutiny.

We also find related implications when considering democratization in this context. Primary barriers to entry described within the communities we reviewed include lack of resources (financial or temporal), musicianship, and disability. Some of these feel distinctly speculative in that they represent barriers to the creation only of \textit{particular} imagined musics, suggesting a model in which democratization is not driven by impediments but a alleged desire to manifest the imagination or aspects of self as music. In such a model, any personality becomes the theoretical \textit{Gesamtkunstwerk}\footnote{The term translates to 'total artwork,’ often used by Wagner to refer to an interdisciplinary artistic practice which expresses a singular voice through all its forms.} if technology can be correctly leveraged. The notion that people represent potential artistic outcomes simply by existing follows an emphasis on individualization, and may also align with \textit{longtermist} perspectives which have become prevalent among AI enthusiasts \cite{ferri2023risk}. In such a philosophy, the potential of a person is equal to or greater than the immediacy of their being, and in our context, we may start to think of the cultural potential dormant in any mind in place of a calculus of future lives.

Shulman stated: “humans are really wired to want to make music” \cite{wilson2024AI}, and the notion that musical creativity is a \textit{dormant ability} is purportedly exposed as music composition is ‘made easy.’ We can read this comment as part of a techno–liberal discourse that fetishises technology to sidestep the contradictions of liberalism, emphasising individualisation while eroding personhood, and bringing this circumvention into contact with the dynamics of networked society \cite[pp.~107–128]{fish2017technoliberalism}.

This perceived \textit{dormant ability} is translated “from [...] mind to music" \cite{suno2024instrumentals} primarily through language, and the ways in which non–musicians can describe music become key. Music in some ways naturally resists language, and this non–discursive and non–propositional character makes it very difficult to describe accurately and succinctly. Text–to–music systems like those examined in this study must be reductive if they are to be controllable \cite{morreale2025reductive}, and we found genre was one of the main ways users described music when interfacing with the systems, being a concise musical descriptor familiar to both musicians and non–musicians. As software is both a product and producer of the world \cite{kitchin2011code}, we see that this means of interfacing with music creation colors how users think about composition, both in replicating the familiar and the potential they see in the systems to be musically innovative. Novelty is used as a synonym for creativity in music \cite{masu2023nime},\footnote{In fact this kind of “social novelty” is also referred to as “historical creativity,” so there is some merit to this confluence \cite[p.~138]{TigreMouraFrancisco2021AibB}.} and as genre is one of the easiest ways to discuss musical forms which have been \textit{established}, ideas of breaking with that established practice tend to express themselves as genre hybridity. This take can be seen in this collection of Suno user quotes:
\begin{quote}
    Wanted to check the creativity of SUNO by making a fusion song…; I'm trying to create a new genre of music; I love the idea of making new genres[...] using AI tools,[...] ones nearly impossible with regular instruments; I'd love to see a completely new genre of music come out of this platform that becomes popular.
\end{quote}
This may also be influenced by what users see explored in other creative mediums utilizing AI, i.e., image generation, where novel combinations are an easy means of demonstrating the systems capabilities. We can understand these issues as techno–solutionist \textit{spandrels}\footnote{As in biology, not architecture.} in this ‘emancipatory’ technology, the reduction of music via language as a consequence of the growth of participation.

\section{Conclusion}
The ways in which companies and developers rhetorize Generative AI in music impacts user experience, and reveals the development of a shared ideology, both \textit{between} companies and their user bases, and \textit{across} the public–facing field.\\
We identified the development of an individualist ideology operating within a globalist worldview. Readily understood against a techno–liberal backdrop, democratization is utilized as part of an indirect coercive market force under processes of governmentality, involving the development of a Marxian “false consciousness” to keep value and function vague but friendly, evading scrutiny. Anthropomorphized and naturalized, the idea of ‘AI assistance’ is used to render developing AI technologies a benign but compelling musical force.

This ‘total ideology’ positions AI along an inevitable flow of innovations in music, against which democratization becomes a noble and necessary means of engagement. We find a link between rhetoric and reported user experience, and the corporate conception of a dormant creative potential appears to be mirrored in community dialogues. Curation is highlighted as a compositional practice, appealing to an individualist framing in which independent musical ecosystems might emerge around a single creator–consumer.

The elevation of language in text–to–music systems reaffirms the centrality of genre, and in requiring music to take a discursive form, potentially inhibits it. However, the ideological construction of friendly, user–focused augmented AI may not be entirely insidious, and the study finds that users reclaim something of the system, and reform shared agency to suit their needs. Such acts of reclamation raise further questions about the location of agency and creativity in these systems, to be explored in future research.

Future studies would benefit from expanding the scope of data gathered by undertaking a more longitudinal study. Although the temporal limitations of this study were necessary due the practical scope, it does limit the outcomes due to the accelerated development of the field. However this study still holds value as the ideological elements we discuss within commercial AI–music are foundational, and have implications for future actions. Additionally as AI Music Studies is still emerging, this paper adds to a growing account of the field, and a historical snapshot of a significant moment.

\section{Authors' contributions}
LP conceptualized the work, lead all phases of the study, and wrote the original draft. FM was involved in supervisory and writing (review and editing) capacities.

\section{Competing Interests}
The authors have no competing interests to declare.

\printbibliography

\end{document}